\documentclass[twocolumn,showpacs,preprintnumbers,
               amsmath,amssymb,floatfix,10pt]{revtex4}
\usepackage{graphicx}
\usepackage{dcolumn}
\usepackage{bm}
\usepackage{graphicx}
\usepackage{subfigure}
\usepackage{amsmath}
\usepackage{amssymb}
\usepackage{wrapfig}
\usepackage{color}
\usepackage[english]{babel}
\usepackage[warning,math]{easyeqn}
\usepackage[thinlines]{easybmat}

\newcommand{\scs}{\scriptscriptstyle}

\def\<{\langle}
\def\>{\rangle}
\begin{document}
%
\preprint{}
%
\title{Energy transfer in nonlinear network models of proteins}
\author{F. Piazza} 
\affiliation{Ecole Polytechnique F\'ed\'erale de Lausanne,
Laboratoire de Biophysique Statistique, ITP--SB,  
BSP-720, CH-1015 Lausanne, Switzerland}  
\author{Y.--H. Sanejouand}
\affiliation{
Laboratoire de Biotechnologie, Biocatalyse et Bior\'egulation, 
UMR 6204 du CNRS,
Facult\'e des Sciences et des Techniques,
2, rue de la Houssini\`ere, 44322 Nantes Cedex 3, France}
%
\begin{abstract} 
We investigate how nonlinearity and topological disorder affect the energy
relaxation of local kicks in coarse-grained network models of proteins.
We find that nonlinearity promotes long-range, coherent transfer of substantial
energy to specific, functional sites, while depressing transfer to generic locations.
Remarkably, transfer can be mediated by the self-localization of discrete breathers
at distant locations from the kick, acting as efficient energy-accumulating centers.
\end{abstract} 
%
%
\pacs{87.14.E-, 87.15.A-, 63.20.Pw}
%
%
%
%
%
%
%
\maketitle
%

\noindent It is now well established that the functional dynamics of proteins 
is deeply rooted in the peculiar topological arrangement of their
native folds, as revealed by many experimental and computational 
studies~\cite{Best:2007qe,Torchia:2003oq}.
In particular, the success of coarse-grained elastic network models (ENMs) 
in describing atomic fluctuations at room temperature
have helped elucidate, at the harmonic level, the subtle interplay between structure and dynamics
on one side and biological function on the 
other~\cite{Tirion:1996mz,Bahar:97,Hinsen:98,Tama:01,Maritan:02,Delarue:02,
Gerstein:02,Phillips:07}.
\\
\indent However, protein dynamics is strongly 
anharmonic~\cite{Xie:2000fk,Edler:2004uq}, a property which has 
to be taken into account in order to rationalize crucial biological processes such as 
energy storage and transfer upon ligand binding, chemical reaction,  {\it
etc}~\cite{Straub:00,Leitner:2008rw}. Yet, even though many theoretical studies
suggest that nonlinear excitations may play an active role in
protein functioning~\cite{Aubry:01,Archilla:2002lr},
the rich phenomenology residing in the interplay between 
protein topology and nonlinearity still remains widely unexplored.
Along these lines, we have recently introduced the Nonlinear Network Model (NNM),
showing how known nonlinear effects can be modulated by the underlying 
non-regular topology of protein systems. For instance, within a large collection of 
enzyme structures, the formation of localized, robust
nonlinear modes appears strongly favored at few specific sites, that 
often lie in close proximity of known catalytic sites~\cite{Juanico:2007ez,Piazza:2008ug}.
\\
\indent In this paper we examine the effects of the nonlinearity/topology interplay on 
energy transfer phenomena across protein structures. Within the NNM framework
a protein is represented by $N$ fictive particles (amino acids) of identical 
mass $M = 110 $ a.m.u., at equilibrium at the corresponding C$_{\alpha}$ site as 
specified in the experimentally determined structures (X-ray or NMR). By imposing 
a fixed cutoff $R_{c}$ on the latter set of coordinates, a protein is
mapped onto a network of nonlinear oscillators, whose potential energy reads
\begin{equation}
\label{FPU}
  U = \sum_{i=1}^N u_{i} \stackrel{\rm def}{=} \sum_{i=1}^N 
                                                \left[ 
                                                    \sum_{j=1}^N  c_{ij}
                                                    \sum_{p=2,4} \frac{k_{p}}{2p}  (r_{ij}-R_{ij})^p
                                                \right]
\end{equation}
where $r_{ij}$ is the distance between residues $i$ and $j$,  $R_{ij}$ their distance in the equilibrium
structure  and $c_{ij} = \{ 1 \,\, {\rm if} \,\, R_{ij} \leq R_{c}, 0 \,\,Ê{\rm otherwise} \}$ is the 
connectivity matrix.
As in previous studies~\cite{Juanico:2007ez}, we take $R_c=$10~\AA, $k_{4} = 5$ kcal/mol/\AA$^4$ and  
fix $k_{2}$ so that the low-frequency part of the linear spectrum match 
actual protein frequencies,
as calculated through realistic force fields~\cite{Brooks:85,Marques:95,Perahia:95}. 
This gives $k_{2}= 5$ kcal/mol/\AA$^2$. The case $k_{4} = 0$
corresponds to the Anisotropic Network Model (ANM)~\cite{Tirion:1996mz,Bahar:97,Hinsen:98}.
\\
\indent Our aim is to investigate how energy initially imparted at a specified site $i$
redistributes across a given structure. To do this, we perform microcanonical
simulations with all residues initially at rest at their equilibrium position
but for a kinetic energy kick at site $i$ of magnitude $E_{0}$. 
Sites in a 3$D$ protein network are not equivalent, featuring e.g. varying connectivity,
clustering coefficient and bond directions. Thus, in order to allow for a comparison of 
energy relaxation from all sites in  a given  structure, the initial 
kick direction ought to be specified by a unique protocol. 
We chose to calculate the directions
of the initial velocities $\vec{v}_{i}(0)$ through the Sequential 
Maximum Strain (SMS) algorithm~\cite{Piazza:2008ug}, which provides an unbiased
measure of the maximum-strain direction at site $i$ for a fixed displacement (here 1 \AA), 
$\hat{e}_{\scs SMS}$.
During the simulation, we record at regular intervals $t_{k}$, $k=1, 2, \dots, N_{s}$ the 
site that carries the highest energy, $n_{k}$, and the value of the latter, 
$e_{n_{k}} = Mv_{n_{k}}^2/2+u_{n_{k}}$.
Corresponding to a fixed simulation time (about 500 ps), we define a transfer probability
from site $i$ (the kicked one) to site $j$ and the fraction of energy transferred as
\begin{equation}
\label{e:Ptr}
\mathcal{P}_{i\rightarrow j} = \frac{1}{N_{s}} \sum_{k=1}^{N_{s}} \delta_{n_{k},j}, \quad
f^{e}_{i\rightarrow j} = \frac{1}{N_{s} \mathcal{P}_{i\rightarrow j}} 
                          \sum_{k=1}^{N_{s}}  \frac{e_j \delta_{n_{k},j}}{E_{0}}
\end{equation}
%
\begin{figure}[ht!] 
\includegraphics[width=\columnwidth,clip]{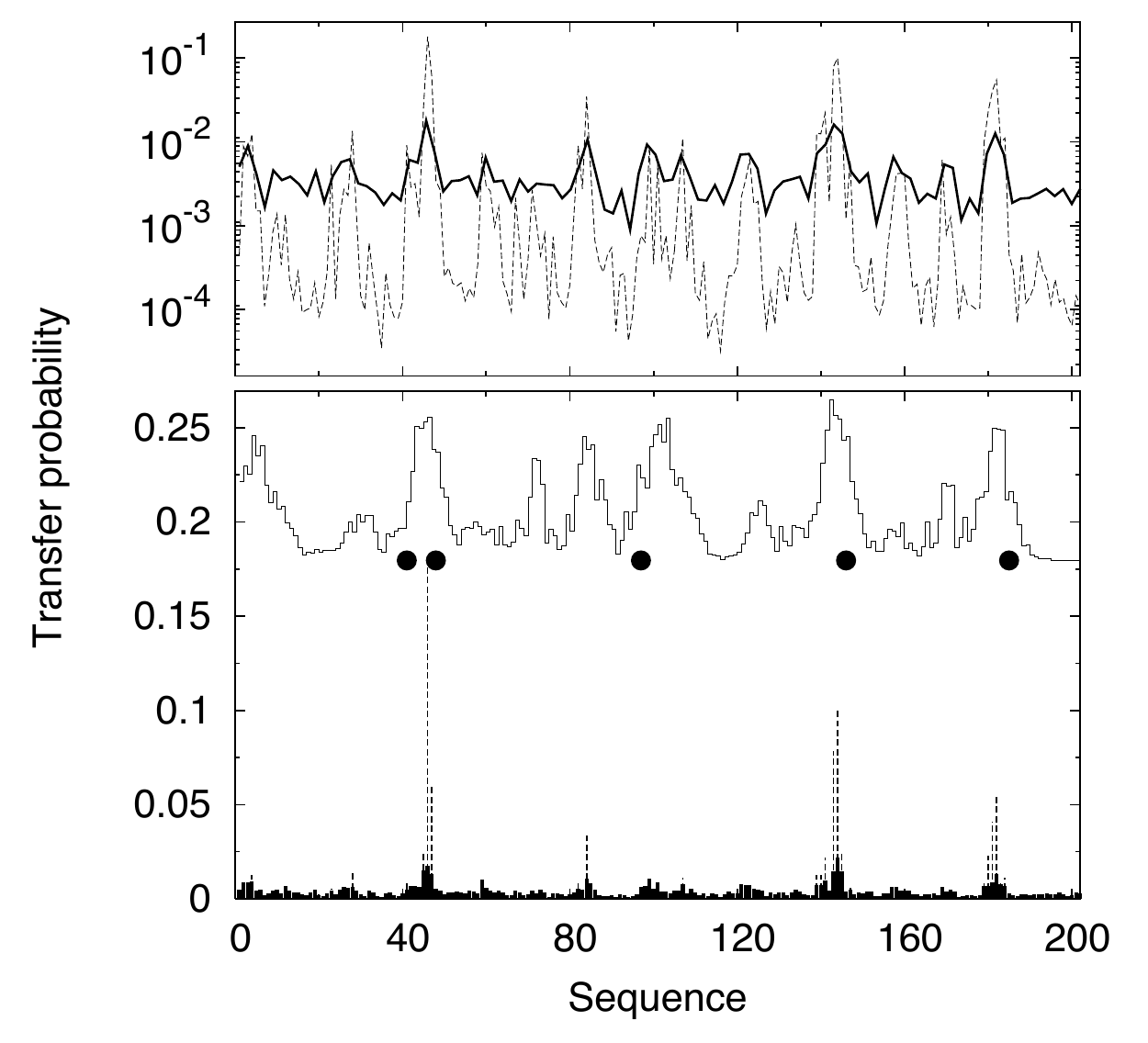}
\caption{Average transfer probability in Riboflavin Synthase (PDB id. 1KZL, $N=202$) on
logarithmic (upper panel) and linear (lower panel) scale. 
Thick solid line: ANM. Dashed line: NNM. The staircase plot in the lower
panel reproduces the stiffness pattern (arb. units). Filled circles flag catalytic sites.
$E_{0}=75$ kcal/mol.\label{f:fig1}}
\end{figure}
\begin{figure*}[ht!]
\centering
\begin{tabular}{c c }
\resizebox{65 mm}{!}{\includegraphics[clip]{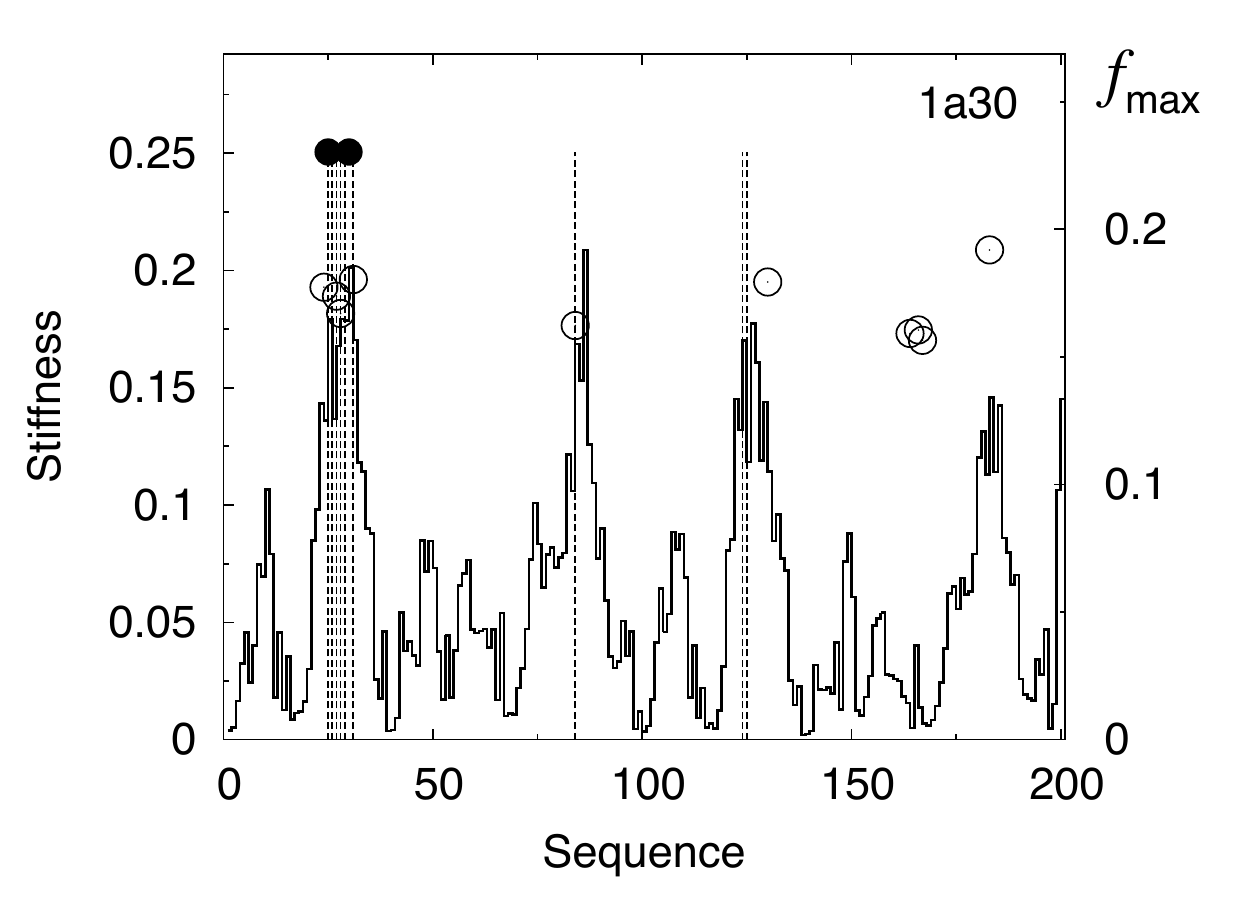}} &
\resizebox{65 mm}{!}{\includegraphics[clip]{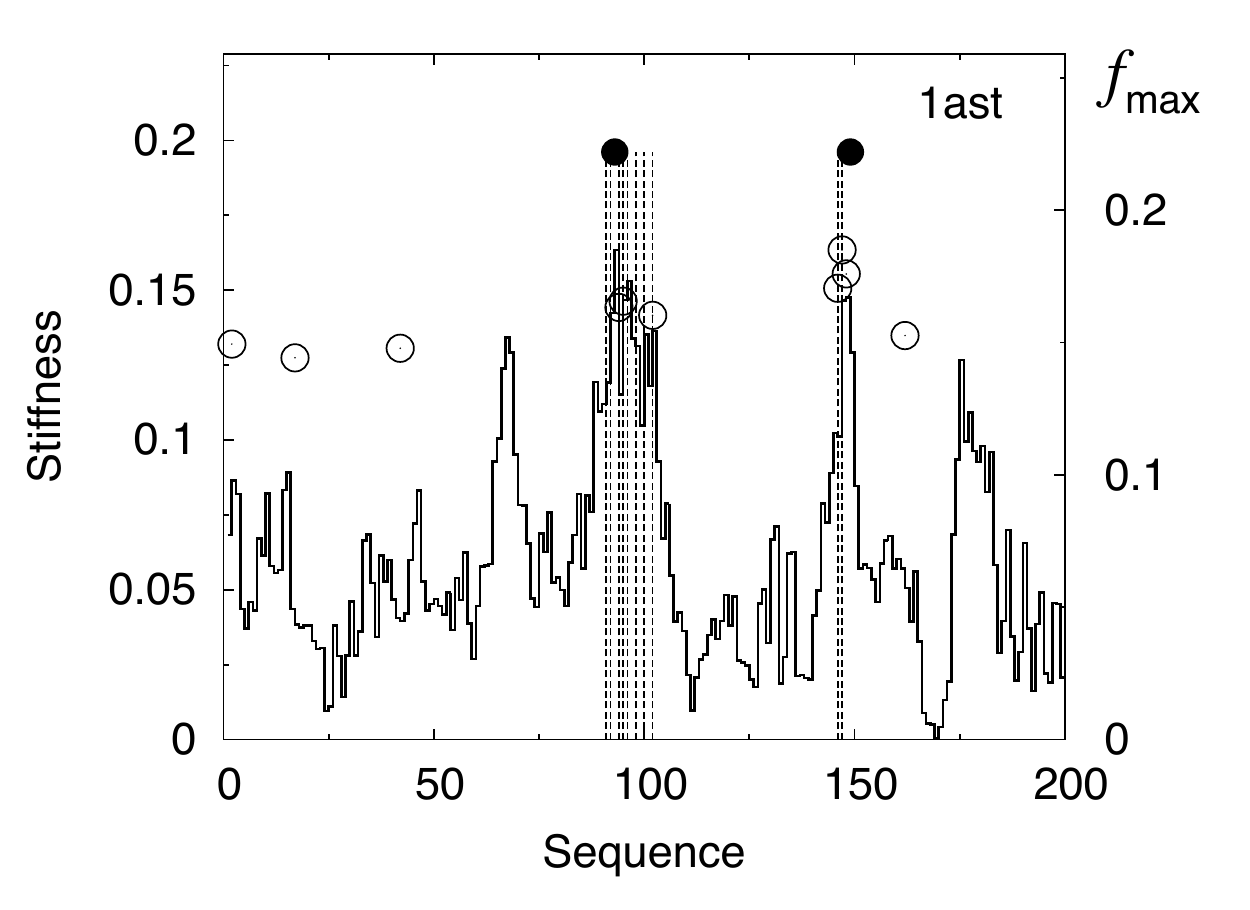}} \\
\resizebox{65 mm}{!}{\includegraphics[clip]{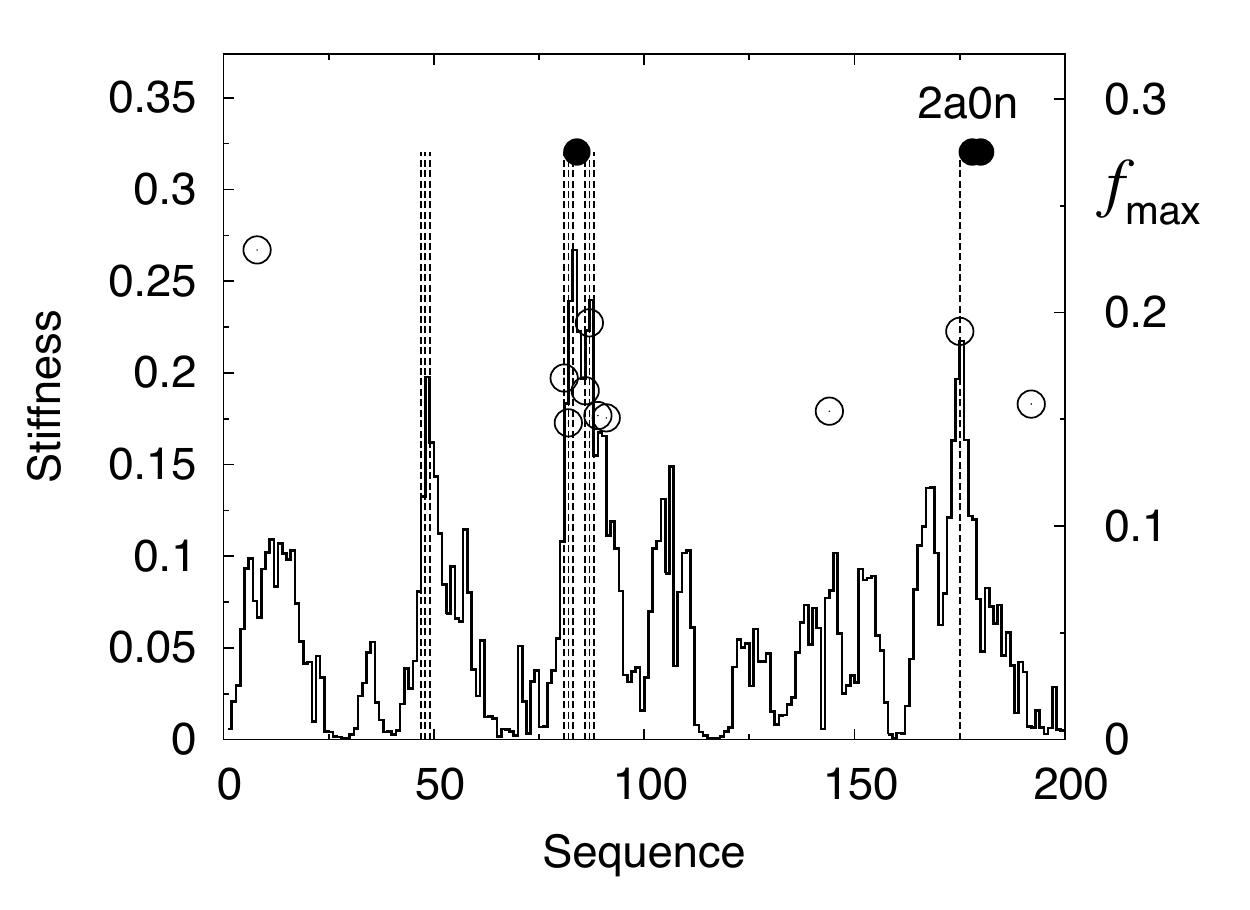}} &
\resizebox{65 mm}{!}{\includegraphics[clip]{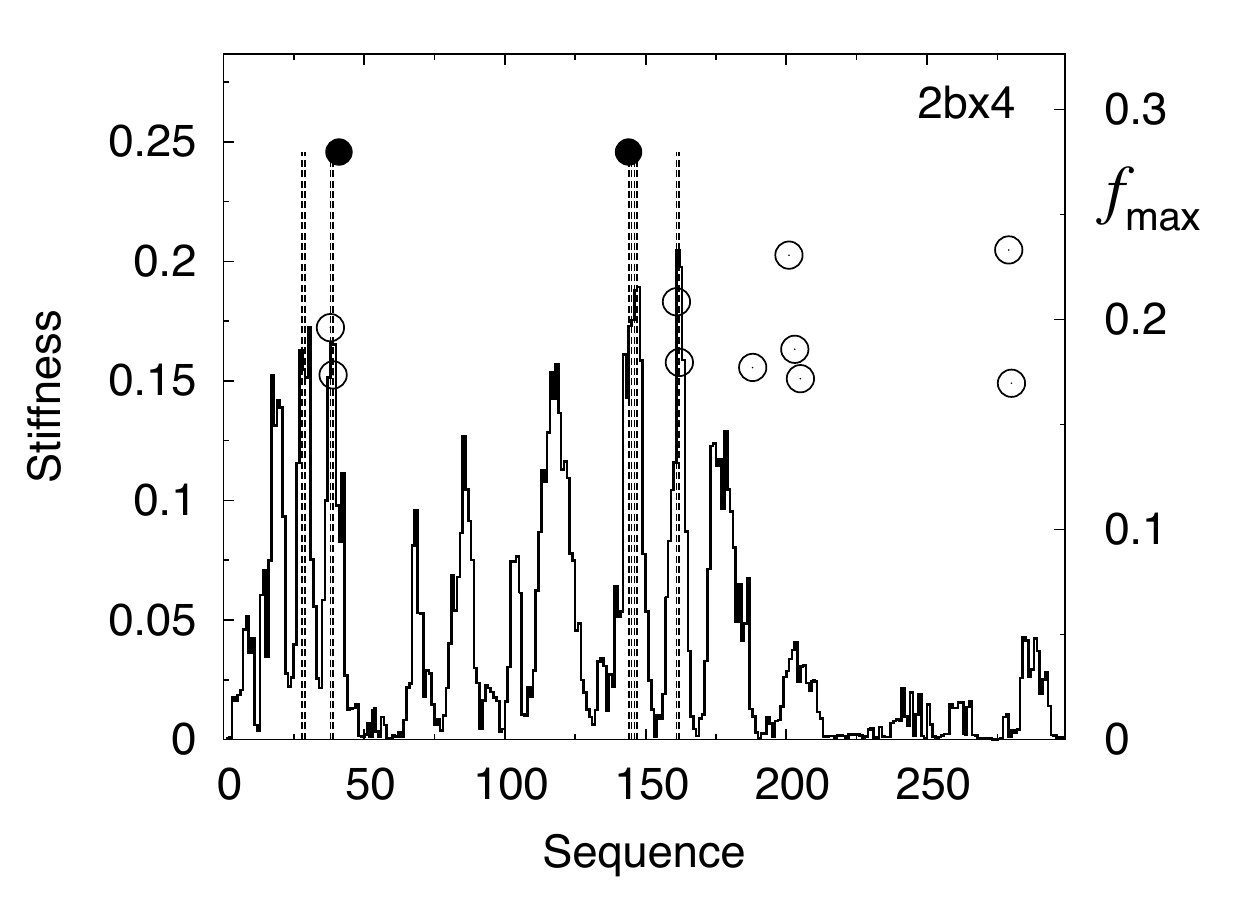}} 
\end{tabular}
\caption{Stiffness plots (staircases) and first ten  sites in the 
ranked list of average energies delivered to 6 \AA-balls around each site (dashed impulses)
for four enzymes: 
HIV-I protease                                     (PDB id. 1A30, $N=201$),
Astacin                                            (PDB id. 1AST, $N=200$),
Imidazole glycerol phosphate synthase subunit hisF (PDB id. 2A0N, $N=200$),
SARS coronavirus main proteinase                   (PDB id. 2BX4, $N=299$).
Filled circles flag catalytic sites. Right axes report the ten largest fractions of energy 
transferred (in units of $E_{0}$), averaged over 6 \AA \ balls,  
versus target ball centers (empty circles). 
$E_{0}=75$ kcal/mol.
\label{f:fig2}}
\end{figure*}
\\
\noindent The first striking result comes from the calculation of {\em average} transfer probabilities.
These gauge  the mean transfer to a given site from kicks at all other sites,
$\< \mathcal{P}_{j} \> = \sum_{i \neq j} \mathcal{P}_{i\rightarrow j} /(N-1)$, obtained from
$N$ independent simulations.
A typical probability transfer plot is shown in Fig.~\ref{f:fig1}. The first notable feature
is that the effect of nonlinearity is to substantially increase the probability of energy funneling to
a few selected sites, while depressing transfer to all other locations with respect to 
the harmonic (ANM) case. Remarkably, the preferred target sites lie in close proximity to 
the known catalytic sites, within the stiffest 
regions~\footnote{We measure the local stiffness $s_{i}$
as a sum over the set $\mathcal{S}$ of ten highest normal modes $\vec{\xi}^k$, 
that is the eigenvectors of the Hessian of the ANM total potential energy, $s_i=
\sum_{j,\alpha} \sum_{k \in \mathcal{S}} c_{ij} [\xi_{j\alpha}^k]^2 / \mathcal{N}_i$,
with $\mathcal{N}_i = \sum_{j} c_{ij}$~\cite{Juanico:2007ez}.}.
Thus, topology and nonlinearity 
team in this case together to sharpen energy funneling to specific functional regions.
\\
\indent The case shown in Fig.~\ref{f:fig1} is not a singular one. In Fig.~\ref{f:fig2} we show the 
stiffness patterns for four other enzymes along with the sites ranking first to tenth as to the 
energy delivered on average to spherical shells with 6 \AA \ radius around each site.
For residue $j$, this amounts to further averaging
the mean energy deposited at sites within the $j$-th ball $\mathcal{B}(j)$, i.e. 
$\<\< f^{e}_{j} \>\> = \< \sum_{i \neq j} f^e_{i\rightarrow j} /(N-1) \>_{\mathcal{B}(j)}$.
%
%
It is manifest that the sites around which most of the energy is deposited invariably spotlight the stiffest regions, 
at the same time identifying functionally relevant locations (see catalytic sites). 
Moreover, the same locations clearly attract substantial fractions of the initial excitation energy, 
as revealed by surveying the maximum transferred energies to each ball $\mathcal{B}(j)$, that is
$f_{\rm max}(j)  = \< \max_{i \neq j} f^e_{i\rightarrow j} \>_{\mathcal{B}(j)}$ 
(empty circles).
%
%
Many events featuring transfers of energy fractions in the range 20 to 25 \% were 
indeed observed.
%
\\
\indent We can learn more on the mechanisms underlying the energy transfer process by examining 
in detail the outcome of a single kick. Fig.~\ref{f:fig3} pictures a long-range transfer 
event occurring when kicking at site LEU 42 in the enzyme Subtilisin. The middle-lower panel (a) shows 
a plot of the most energetic site $n_{k}$ as a function of time, clearly illustrating the transfer
to site VAL 177, some 23 \AA \ away, occurring at $t_{\ast} \approx 275$ ps. The transfer process also involves 
site ALA 85, as a {\em passage} site. Remarkably, a plot of the energy $e_{\rm max}(t)$
of  the most energetic site at time $t$ clearly shows that such passage coincides with a 
redistribution of energy across the structure (see middle panel (c)). Subsequently, energy is 
garnered from the neighborhood and stabilized in a localized mode centered at VAL 177, finally carrying about 20 \% 
of the total energy. This marks the true transfer event.
Such energy-harvesting, self-localized vibrations are  generic in discrete non-linear systems
and are well-known as Discrete Breathers (DB)~\cite{Flach:1998fj}. These are robust, time-periodic 
exponentially localized modes, 
whose vibrational frequency lies outside the linear spectrum of the system. 
In the context of the NNM, we have shown how accurate approximations of such periodic orbits can be calculated analytically,
reproducing the marked affinity of DB self-localization in topologically disordered media 
for the stiffest spots~\cite{Juanico:2007ez,Piazza:2008ug}.
Here we have shown that DBs may also be excited as a consequence of localized impulses at considerable distances from 
the excitation, playing the role of energy-accumulating transfer vectors.
In order to substantiate the above interpretation, we have performed Principal Component Analysis (PCA) on 
an extended portion of the post-transfer dynamics. The power spectrum of the system trajectory projected 
on the first principal mode (PM1) is shown in the upper left panel  of Fig.~\ref{f:fig3}, clearly revealing 
the existence of a nonlinear, time-periodic excitation, reminiscent of chaotic DBs that self-localize
through modulational instability in nonlinear lattices~\cite{Cretegny:1998to}.
\\
\indent More insight as to why energy is transferred from LEU 42 a long distance away can be obtained 
by turning to the space of Normal Modes (NM) $\vec{\xi}^k$ $(k=1,2,\dots,3N-6)$.
Middle panel (b) of Fig.~\ref{f:fig3} reports the mode carrying
the highest energy 
$\varepsilon_{k}(t) = (\dot{Q}_{k}^2 + \omega_{k}^2 Q_{k}^2)/2$ 
at time $t$, where $Q_{k} = \sum_{j,\alpha} x_{j\alpha} \xi^k_{j\alpha} $ is the NM-transform
of the system coordinates $x_{j\alpha}$ ($j=1,2,\dots,N,\alpha=x,y,z$) 
and $\omega_{k}$ are the NM frequencies.  Before the transfer event,
energy is bounced among four high-frequency modes, NMs 1,3,8 and 9. This can be understood 
by constructing the {\em NM overlap network} starting from the NMs with the largest projections on the 
initial excitation unit vector $\hat{e}_{\scs SMS}$ are greatest (four modes in the first column of the 
bottom graph in Fig.~\ref{f:fig3}, NMs 138,127,67 and 37, making up about 60 \% of $\hat{e}_{\scs SMS}$). 
For a given NM $p$,  two links are drawn to the two-highest ranking NMs in the ordered list
of absolute overlap coefficients $t_{pq} = \sum_{j,\alpha} |\xi^p_{j\alpha}||\xi^q_{j\alpha}|$.
By doing this for the four NMs involved in the SMS vector,
a closed network emerges identifying the NM3 $\substack{\rightarrow \\ \leftarrow}$ 
NM8 $\substack{\rightarrow \\ \leftarrow}$ NM9 
loop. Thus, in the presence of nonlinearity energy is immediately directed to a reduced group of NMs
via resonant overlap mechanisms. This finding agrees with results of atomistic simulations highlighting 
the importance  of spatial overlap for NM-NM energy transfer~\cite{moritsugu}.
\\
\indent High-frequency NMs are strongly localized in space. In particular, 
ALA 85 is the NM site (the site with largest displacement) in NM3 and the second NM site in NM8,
which explains the role of ALA 85 in the energy circulation process. 
Before transfer, however, energy also bounces back and forth from NM1, the highest-frequency mode,
reflecting the nonlinear frequency shift on NM3 toward greater frequencies (see again panel b). 
At $t=t_{\ast}$ energy starts departing the region around LEU 42 and a fluctuation pumping up 
NM3 occurs (panel d), shifting its frequency  upwards by virtue of nonlinearity. The energy at stake is
sufficient to trigger nonlinear localization and a DB finally installs at VAL 177, the NM site of NM1,
gathering vibrational energy from the background. Correspondingly, the energy on NM1 increases (see panel d). 
To substantiate the above analysis, we have calculated analytically
the DB mode pattern centered at site VAL 177 
with the technique described in Ref.~\onlinecite{Piazza:2008ug}. Then we have built the 
network connecting the first two principal modes, the first three NMs and the DB, where the links
are weighted by the normalized scalar products (upper graph in Fig.~\ref{f:fig3}). As it shows, the 
PMs essentially reflect the underlying competition between NM1 and NM3. In particular, 
the first principal mode confirms the excitation of a DB emerging as a nonlinear continuation 
of the edge normal mode, as predicted theoretically in  Ref.~\onlinecite{Piazza:2008ug}.
In agreement with this picture, kicks at ALA 42 of weaker energy resulted in a DB installing at MET 199,
the NM site of NM2. That is, less energy causes a smaller frequency shift and the DB branch
originating from the continuation of NM2 is excited instead. Reducing $E_{0}$ further, the transfer 
is observed to halt at ALA 85, as explained by the NM overlap network.   
\\
\indent In this paper we have shown how nonlinearity in a topologically non-regular system 
boosts energy transfer to few specific locations. In enzyme structures, these coincide invariably 
with the stiffest regions, also hosting the functionally relevant sites. Nonlinearity sharpens 
the transfer selectivity, by reducing at the same time the transfer probability to generic locations.
The energy transferred by virtue of nonlinearity may be a conspicuous portion of the initial excitation,
in which cases localized vibrations akin to Discrete Breathers self-localize as energy-collecting centers,
often realizing amazingly efficient energy transfer channels across considerable distances.

\begin{figure*}[t!]
\centering
\resizebox{15 cm}{!}{\includegraphics[clip]{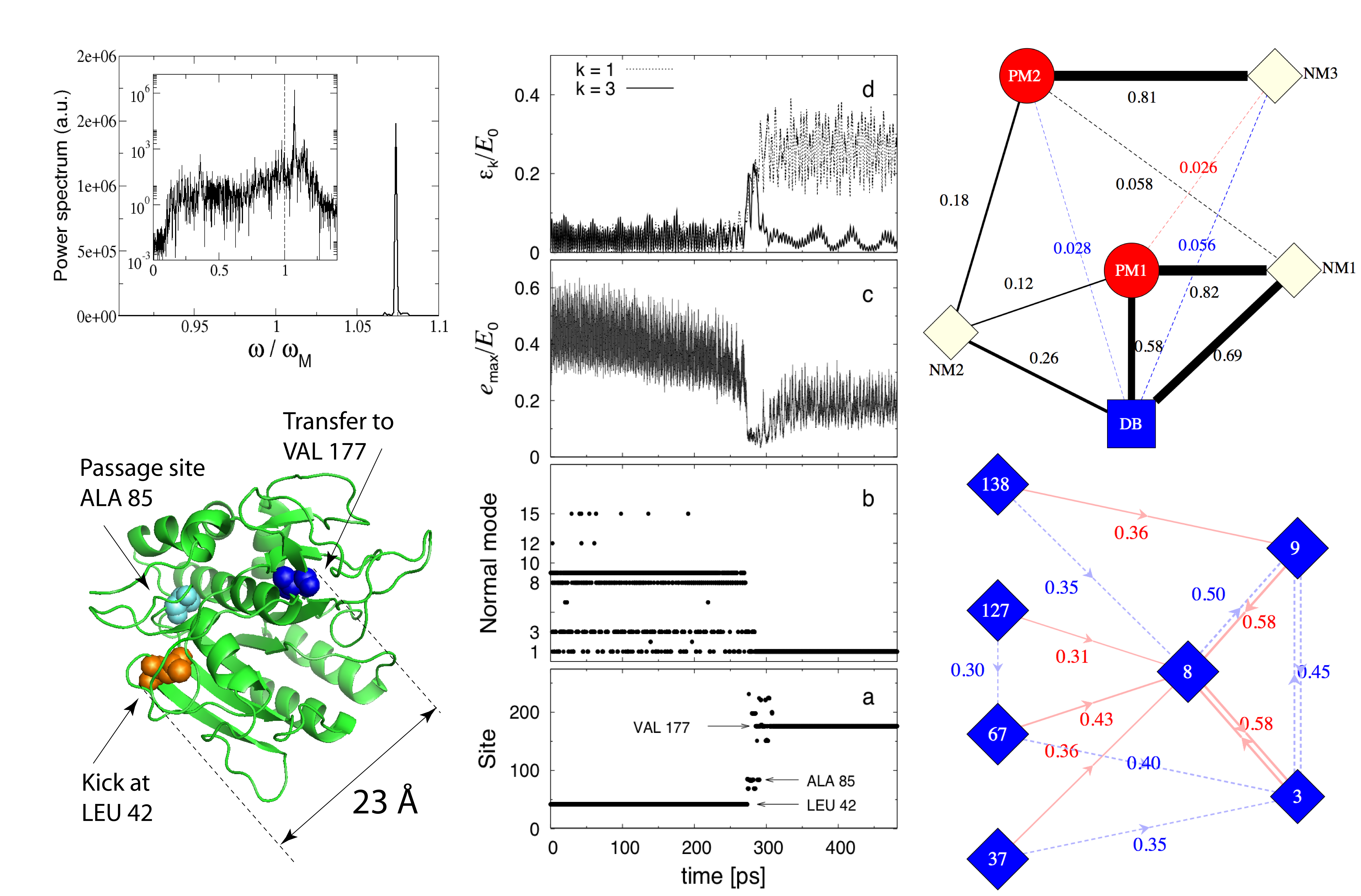}}
\caption{(Color online) Kick at site LEU 42 in Subtilisin (PDB id. 1AV7, $N=274$). Middle panels:
most energetic site vs time (a), most energetic normal mode vs time -  NMs being ranked in order  
of decreasing eigenfrequency -  (b), 
highest site energy vs time (c),
energies of two NMs vs time (d). Upper left panel: power spectrum of the system trajectory 
for $t > t_{\ast}=275 $ ps  projected on the first principal mode;
$\omega_{\scs M} = 101.2$ cm$^{-1}$ is the 
band-edge linear frequency. Lower right graph: NM overlap network. Nodes are 
NMs, red and blue links connect to nodes ranking first and second, respectively, in overlap (see text). Link
weights are the overlap coefficients $t_{pq}$. Upper right graph. Network relating principal modes (PM), NMs and
the analytical Discrete Breather pattern (DB). Link weights are the absolute cosines
(normalized scalar products). In both graphs 
the link width is proportional to its weight. $E_{0}=100$ kcal/mol. 
}
\label{f:fig3}
\end{figure*}

%
%

\end{document}